\documentstyle[aps,preprint,epsf]{revtex}

\begin {document}
\title
{PROTON STRUCTURE IN TRANSVERSE SPACE AND THE EFFECTIVE CROSS SECTION}

\author{G. Calucci and D.Treleani} 

\address{ Universit\`a di Trieste; Dipartimento di Fisica Teorica, 
Strada Costiera 11, Miramare-Grignano, and
INFN, Sezione di Trieste, I-34014 Trieste, Italy.}

\maketitle

\begin{abstract}
The observation of double parton collisions by CDF has provided the first
direct information on the structure of the proton in transverse space.
The actual quantity which has been measured is the `effective cross section'
$\sigma_{eff}$,
which is related to the transverse size of the region where hard
interactions are localized. The actual value which has been measured 
is sizably smaller than naively expected and it is 
an indication of important correlation effects in the many-body parton
distribution of the proton. We discuss the problem pointing out a 
possible source of correlations in the proton structure, which could 
have a significant effect on the value of $\sigma_{eff}$.
\end{abstract}
\vspace{3cm}

E-mail giorgio@trieste.infn.it \\

E-mail daniel@trieste.infn.it \\

\newpage

\section{Introduction}
While parton distributions represent the relevant information on the non 
perturbative input
for most of the large $p_t$ processes at present energy, parton distributions 
do not exhaust the information on the hadron structure. In fact the information
in the parton distributions corresponds to the average number of partons of a
given kind and with a given momentum fraction $x$ which is seen when probing
the hadron with the
resolution $Q^2$. In this respect a qualitative change has occurred when it has
been shown that the observation of hadronic interactions with
multiple parton collisions is experimentally feasible. 
In a multiparton collision event different pairs
of partons interact independently at different points in the transverse
plane\cite{dpth} and the process, as a consequence,   
depends in a direct way on the actual 
distribution of the interacting partonic matter in 
transverse space. The simplest case is the double parton 
scattering. The non
perturbative input to the process is the double parton distribution 
$D_2(x,x'; {\bf b})$ depending on the momentum fractions of the interacting 
partons and on their relative distance in the transverse plane ${\bf b}$. 
If all partons are uncorrelated and the dependence on the different 
degrees of freedom is factorized, one may write 
$D_2(x,x'; {\bf b})=g_{eff}(x)g_{eff}(x')F({\bf b})$ where $g_{eff}(x)$ is the 
usual effective 
parton distribution, namely the gluon plus 4/9 of the quark and anti-quark
parton distributions. When the interacting hadrons are identical $F({\bf b})$ 
is equal to the overlap in transverse plane
of the matter distribution of the two hadrons (normalized to one)
as a function of ${\bf b}$, 
which now represents the impact parameter. 
The effective cross section
$\sigma_{eff}$ is introduced by the 
inclusive double scattering process, which is proportional to
$1/\sigma_{eff}=\int d^2bF^2({\bf b})$\cite{dpth}. 
The effective cross section is then a well
defined property of the hadronic interaction and, at least if the simplest
possibility
for the hadronic structure is realized, it is both energy
and cutoff independent. Analogously the dimensional scale factors,
which one may introduce in relation to triple, quadruple etc.
partonic collisions, are energy and cutoff independent properties of
the interaction.

Initially the search of double parton collisions have been sparse 
and not very
consistent \cite{dpex}. CDF however has recently claimed the observation of a 
large number of double parton scatterings \cite{cdf}. The measured value of 
the effective cross section,
$\sigma_{eff}=14.5\pm1.7^{+1.7}_{-2.3}mb$, is sizably smaller than expected 
naively and it is an
indication of important correlation effects in the hadron
structure \cite{ctlast}. Correlations on the other hand are a manifestation of
the links between the different constituents of the hadronic
bound state and, in fact, 
it is precisely this sort of connections which one would like to learn as a 
result of experiments probing the hadron structure. 

In the present paper we point out a rather natural possible source
of correlations, which is
able to give rise to a sizable reduction of the value of the effective cross
section, as compared to the most naive expectation.
In fact we show that, when linking the transverse size of the 
gluon and sea distributions to the actual configuration of the valence, 
the expected value of $\sigma_{eff}$ is sizably decreased. In the next
paragraph
we recall the main features of the multiple partonic collision processes in the
simplest uncorrelated case, in the following section 
we correlate the distributions of gluons and
sea quarks with the configuration of the valence and, in the final paragraph, we
discuss our results.

\section{Multiple parton collisions}

The non-perturbative input to the multiple parton collision processes is
the many body parton distribution. In most cases the simplest uncorrelated 
case, with
factorized dependence on the actual degrees of freedom, is considered and
the probability
distribution is expressed as a poissonian. The probability of a
configuration with $n$ partons with fractional momenta $x_1,\dots,x_n$ and
with transverse coordinates ${\bf b}_1,\dots,{\bf b}_n$ is then expressed as
\begin{equation}
\Gamma(x_1,{\bf b}_1,\dots x_n,{\bf b}_n)=\frac{D(x_1,{\bf b}_1)
\dots D(x_n,{\bf b}_n)}{ n!}\exp\Big\{-\int D(x,{\bf b})dx d^2b\Big\}
\label{pdistr1}
\end{equation} 
where $D(x,{\bf b})=g_{eff}(x)f({\bf b})$ is the average number of partons
with momentum fraction $x$ and with transverse coordinate ${\bf b}$. 
The function $g_{eff}(x)$ is the effective
parton distribution while $f({\bf b})$ represents the partonic matter 
distribution in transverse space. The many body parton distribution is
an infrared divergent quantity, one needs therefore to introduce an infrared
cut-off. A natural cutoff in this context is the lower value of momentum 
transfer
which allows a scattered parton to be recognized as jet in the final state.
Given the many-body parton distribution
one can work out all possible multi-parton interactions. If rescatterings are
neglected, namely if every parton is allowed to interact with large
momentum transfer (namely larger than the infrared cutoff) at most once, one
may write a simplest analytic expression for the hard cross section
$\sigma_H$, corresponding to the cross section counting all inelastic hadronic
events with al least one hard partonic interaction\cite{ametll}:
\begin{equation}
\sigma_H=\int d^2\beta\Bigl[1-e^{-\sigma_SF(\beta)}\Bigr]
 =\sum_{n=1}^{\infty}\int d^2\beta{\bigl(\sigma_SF(\beta)\bigr)^n\over n!}
 e^{-\sigma_SF(\beta)}
\label{sigh}
\end{equation}
here $\beta$ is the impact parameter of the hadronic collision,
$F(\beta)=\int d^2bf({\bf b})f({\bf b}-\beta)$ and $\sigma_S$ is
the usual expression for the integrated inclusive cross section to produce jets,
namely the convolution of the parton distributions and
of the partonic cross section. The unitarized expression of $\sigma_H$ in
Eq.\ref{sigh}, which takes into account the possibility of many
parton-parton interactions in an overall hadron-hadron collision, 
is well behaved in the infrared region and it corresponds to
a poissonian distribution of multiple parton collisions at a fixed value
of the impact parameter of the hadronic interaction. 
If one works out in Eq.\ref{sigh} the average number of partonic
collisions one obtains 
\begin{equation}
\langle n\rangle\sigma_H=\int d^2\beta\sigma_SF(\beta)=\sigma_S
\end{equation}
so that $\sigma_S$ represents the inclusive cross section normalized with the
multiplicity of partonic interactions.
The inclusive cross section for a double parton scattering $\sigma_D$, 
normalized in an
analogous way with the multiplicity of double parton collisions, $\langle
n(n-1)/2\rangle\sigma_H$, is given by
\begin{equation}
\frac{\langle n(n-1)\rangle}{2}\sigma_H=
\frac{1}{2}\int d^2\beta\sigma_S^2F^2(\beta)=\sigma_D
\end{equation}
One may then introduce the effective cross section:
\begin{equation}
\sigma_D=\frac{1}{2}\frac{\sigma_S^2}{\sigma_{eff}}
\label{sigeff}
\end{equation}
which is therefore expressed in terms of the overlap of the partonic matter
distribution of the two interacting hadrons as
\begin{equation}
\sigma_{eff}=\frac{1}{\int d^2\beta F^2(\beta)}
\end{equation}

\begin{figure}
\centerline{
\epsfysize=8cm \epsfbox{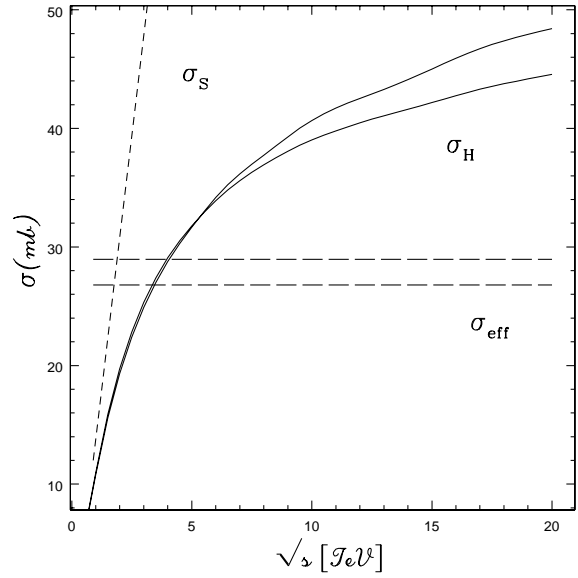}  
}
\vspace{1 cm}
\caption[ ]{ $\sigma_S$, $\sigma_H$ and $\sigma_{eff}$ in the simplest 
uncorrelated case. The two curves for $\sigma_H$ and $\sigma_{eff}$ refer to 
the two different choices made for 
$f({\bf b})$, a gaussian (higher curve) and a sphere (lower curve).}
\label{simplest}
\end{figure}

In fig.\ref{simplest} two different analytic forms for $f({\bf b})$ are
compared: a gaussian and the projection of a sphere in transverse space.
The radius of the distributions has been fixed in such a way that in
both cases the RMS radius has the same value that,
following CDF\cite{cdf}, has been set equal to $0.56$fm. 
For the cutoff we have used $5$GeV and, in order to be able to reproduce
the value of the integrated hard cross section for producing minijets
measured by UA1\cite{UA1}, we have multiplied the partonic cross section,
computed in the perturbative QCD parton model, by the appropriate $k$-factor.  
Both choices give 
a similar qualitative result: the effective cross section is constant with
energy (and cutoff independent) while the `total' hard cross section $\sigma_H$
grows rapidly as a function of the c.m. energy. 
The values
obtained for $\sigma_{eff}$ are however too large, by roughly a factor two, as
compared to the value quoted by experiment. One has to add that in the
experimental analysis of CDF all events with triple parton collisions have been
removed from the sample of inelastic events with double parton scatterings. 
The resulting value for the effective cross section,
$\sigma_{eff}|_{CDF}$, is therefore larger with respect to the quantity 
discussed here above and usually considered in the literature.
The disagreement with the most naive picture is therefore even stronger
then apparent when comparing the result of the uncorrelated calculation with
the quoted value of $\sigma_{eff}$.

In the usually considered picture of multiparton interactions just recalled
all correlations are neglected and one may therefore claim that
the experimental evidence is precisely that correlations play an important 
role in the
many-body parton distribution of the hadron. In the next paragraph
we propose therefore a slight modification to the picture,
linking the transverse size
of the gluon and sea distributions to the configuration in
transverse space of valence quarks.

\section{A simplest model for the partonic matter distribution in the proton}

Charged matter is distributed in the
proton according to the charge form factor, which is well
represented by an exponential expression in coordinate space. 
The information refers to a large
extent to the distribution in space of the valence quarks, which can then be
found in various different configurations in transverse space with a given 
probability distribution. 
Less is known on the distribution in space of the neutral matter component
of the proton structure, namely the gluons. In this respect one may consider
two different extreme possibilities: 
\begin{itemize}
\item the distribution in space of sea quarks and gluons has no relation with
the distribution of valence quarks,
\item or, rather, its transverse size is linked closely to the actual
configuration in space of the valence quarks.
\end{itemize}
The no correlation hypothesis, which has been ruled out by the measure
of $\sigma_{eff}$, would imply the first possibility. 
We try therefore to work out here a simplest model where the distribution of 
the whole partonic matter in the hadron is driven by the actual configuration of
valence quarks.

The parton distribution in Eq.\ref{pdistr1} can be modified by separating the
valence quarks from gluons and sea quarks and by correlating the transverse
size of the distributions of sea quarks and gluons with the transverse size 
of the actual configuration of the valence. If however one simply rescales 
the distribution
function in transverse space $f({\bf b})$ by rescaling the radius while keeping
the normalization constant, one is imposing a conservation constraint in the number of
gluons. The number is in fact the same in each configuration of the
proton, both when it is squeezed to a small transverse size and
when it is expanded to a relatively large dimension. The picture therefore 
is not consistent with the common believe that the 
energy to the gluon field grows basically because of the growth of the 
distance between the valence quarks.
It seems therefore more reasonable to remove the constraint on normalization
and to keep rather fixed the maximum value of $f({\bf b})$ in all
configurations. We modify therefore the simplest poissonian expression
in Eq.\ref{pdistr1} and we express the probability distribution
$\Gamma$ as follows:
\begin{eqnarray}
\Gamma(x_1,{\bf b}_1,\dots x_n,{\bf b}_n)&=&\varphi({\bf
B}_D,{\bf B})q_v(X_1)q_v(X_2)q_v(X_3)\nonumber\\
&\times&\frac{1}{n!}
\bigg[g(x_1)f(b,{\bf b}_1)\frac{b^2}{\langle b^2\rangle}
\dots g(x_n)f(b,{\bf b}_n)\frac{b^2}{\langle b^2\rangle}\bigg]
{\rm exp}\Big\{-\frac{b^2}{\langle b^2\rangle}\int g(x)dx \Big\}
\label{pdistr2}
\end{eqnarray} 
To simplify the notation we have not written explicitly
the dependence of $\Gamma$ on the coordinates of the valence quarks, $X$ and 
${\bf B}$. The dependence of $\Gamma$ on the momentum fraction of
the valence quarks $X$ is factorized and $q_v(X)$ is the usual distribution of
valence quarks as a function of the momentum fraction $X$.
The function $f(b,{\bf b}_i)$ represents the distribution
of gluons and sea quarks in transverse space. It is a function
of the transverse coordinate of the considered parton, ${\bf b}_i$, and
it depends on the scale factor $b$ whose value is given by
the actual configuration of the valence in transverse
space. 
The transverse coordinates of the three valence quarks are 
\begin{eqnarray}
{\bf B}_1&=&\frac{1}{2}{\bf B}_D+{\bf B}\nonumber\\
{\bf B}_2&=&\frac{1}{2}{\bf B}_D-{\bf B}\nonumber\\
{\bf B}_3&=&-{\bf B}_D
\end{eqnarray}
The dependence on the transverse coordinates of the valence is given by
$\varphi({\bf B}_D,{\bf B})$, that is the integral on the longitudinal
coordinates $Z_D,Z$ of $\phi({\bf R}_D,{\bf R})$, representing the 
valence structure of the proton in coordinate space. 
Explicitly we use the exponential form
\begin{equation}
\phi({\bf R}_D,{\bf R})=\frac{\lambda_D^3\lambda^3}{(8\pi)^2}
\exp\big\{-(\lambda_DR_D+\lambda R)\big\} 
\end{equation}
where
\begin{eqnarray}
\lambda_D&=&\frac{2\sqrt{3}}{\sqrt{\langle r^2\rangle}}\nonumber\\
\lambda&=&\frac{4}{\sqrt{\langle r^2\rangle}}
\end{eqnarray}
and $\sqrt{\langle r^2\rangle}=0.81$fm is the proton charge radius.

In a given configuration
of the valence, the average number of gluons and sea quarks is not equal to
the overall average number $g(x)$ (with $g(x)$ we indicate here the
sum of gluon and sea quark distributions). It is rather equal to
$g(x)\frac{b^2}{\langle b^2\rangle}$; where $\langle b^2\rangle$ is the average of
$b^2$ with the probability distribution of the valence. To make a definite
choice we take $b=B_D$ in such a way that
\begin{equation}
\langle b^2\rangle=\langle B_D^2\rangle=\int d{\bf R}_Dd{\bf R}B_D^2\phi({\bf R}_D,{\bf R})
\end{equation}
The density of gluons and sea quarks in transverse space is then
constant in the middle of the proton in all different configurations of the 
valence quarks and it is equal to the value assumed in the average
configuration. The actual number of gluons and sea quarks is therefore a 
function of
the configuration taken by the valence, compact configurations giving rise to
small numbers while more extended configurations giving rise to larger numbers.

Summing over all possible probability configurations of
gluons and sea quarks one obtains, from Eq.\ref{pdistr2}, the probability
distribution of valence in transverse space $\varphi({\bf B}_D,{\bf B})$.
The average number of gluons and sea quarks, with given momentum fraction $x$,
is 
\begin{equation}
\int d{\bf B}_Dd{\bf B}d{\bf b}\varphi({\bf B}_D,{\bf B})
g(x)f(B_D,{\bf b})\frac{B_D^2}{\langle B_D^2\rangle}=g(x)
\end{equation}
while the average number of partonic collisions, non involving 
valence
$\langle n({\bf B}_D,{\bf B},{\bf B}'_D,{\bf B}',\beta)\rangle$, 
for a given configuration of the valence and for a given value of
the impact parameter $\beta$, is expressed as
\begin{eqnarray}
\langle n({\bf
B}_D,{\bf B},{\bf B}'_D,{\bf B}',\beta)\rangle=
\int d{\bf b}f(B_D,{\bf b}-\beta)\frac{B_D^2}{\langle B_D^2\rangle}
f(B'_D,{\bf b})\frac{(B'_D)^2}{\langle B_D^2\rangle}
[\sigma_S(g+q_s,g+q_s)]
\end{eqnarray}
The average number of interactions of valence quark with gluons and sea 
is written in an analogous way.
If one integrates the average number of partonic collisions $\langle n({\bf
B}_D,{\bf B},{\bf B}'_D,{\bf B}',\beta)\rangle$ over all configurations of
the valence quarks of the two hadrons, with the corresponding weights, and on 
the hadronic
impact parameter $\beta$, one obtains the single scattering inclusive cross
section $\sigma_S$:
\begin{eqnarray}
\int d{\bf B}_Dd{\bf B}d{\bf B}'_Dd{\bf B}'d\beta
\varphi({\bf B}_D,{\bf B})\varphi({\bf B}'_D,{\bf B}')
\langle n({\bf
B}_D,{\bf B},{\bf B}'_D,{\bf B}',\beta)\rangle
=\sigma_S(g+q_s,g+q_s)
\end{eqnarray}
An analogous expressions for interactions of valence quarks with gluons
and sea quarks is readily written.

All average quantities are therefore the well known ones of the perturbative
QCD-parton model. The inclusion of
the transverse degrees of freedom in the relations above allows one to 
write down promptly all expressions corresponding to the various multiparton
collision processes. For the double parton scattering case one has:
\begin{eqnarray}
\sigma_D=\frac{1}{2}\int d{\bf B}_Dd{\bf B}d{\bf B}'_Dd{\bf B}'d\beta
\varphi({\bf B}_D,{\bf B})\varphi({\bf B}'_D,{\bf B}')
\Big[\langle n({\bf
B}_D,{\bf B},{\bf B}'_D,{\bf B}',\beta)\rangle\Big]^2
\end{eqnarray}
and for the hard cross section
\begin{eqnarray}
\sigma_H=\int d{\bf B}_Dd{\bf B}d{\bf B}'_Dd{\bf B}'d\beta
\varphi({\bf B}_D,{\bf B})\varphi({\bf B}'_D,{\bf B}')
\Big[1-{\rm exp}\Big\{-\langle n({\bf
B}_D,{\bf B},{\bf B}'_D,{\bf B}',\beta)\rangle\Big\}\Big]
\end{eqnarray}
The effective cross section is obtained, as in the uncorrelated case, from
$\sigma_D$ by using Eq.\ref{sigeff}.

\section{Discussion}

\begin{figure}
\centerline{
\epsfysize=8cm \epsfbox{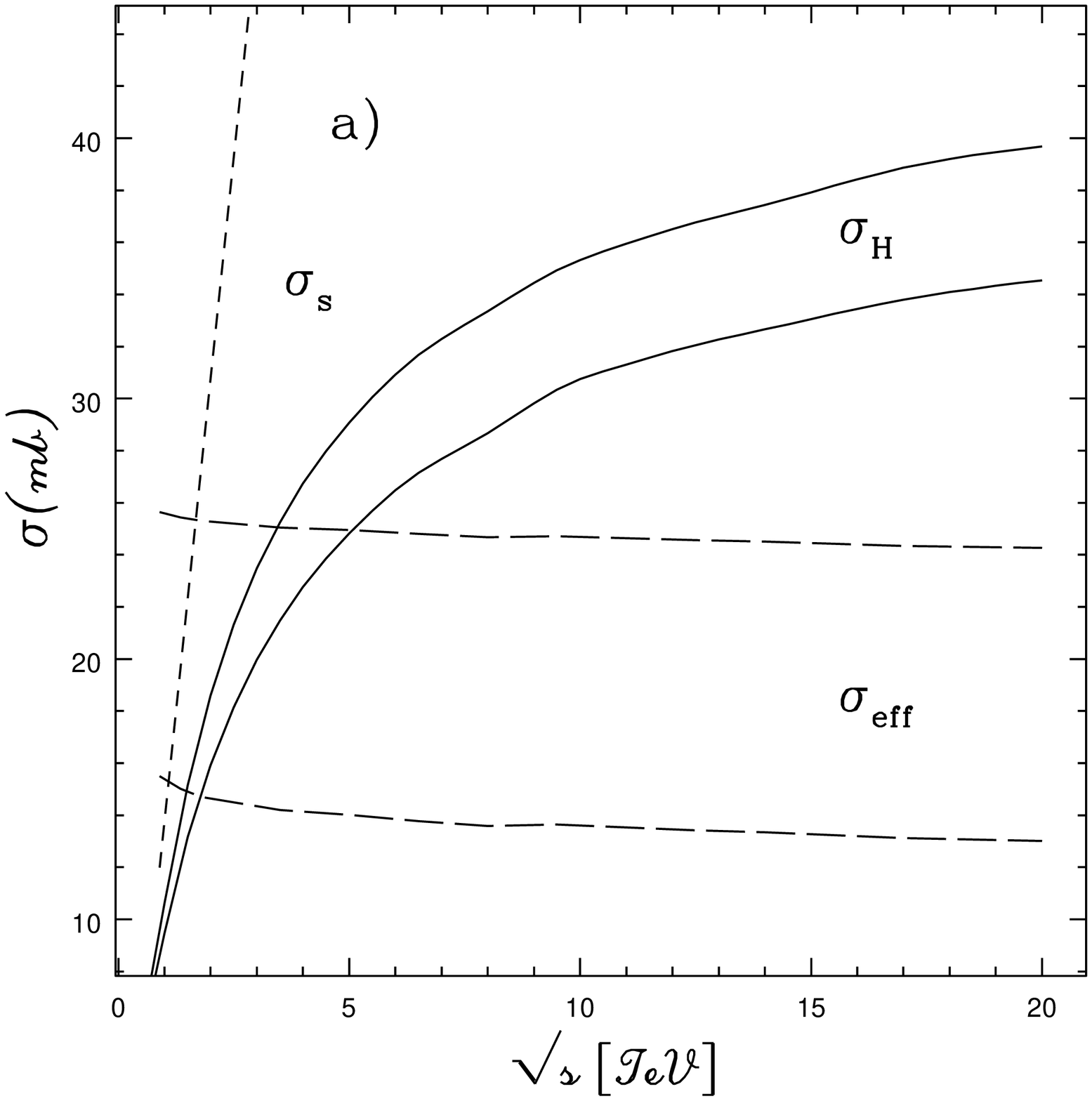}  
\epsfysize=8cm \epsfbox{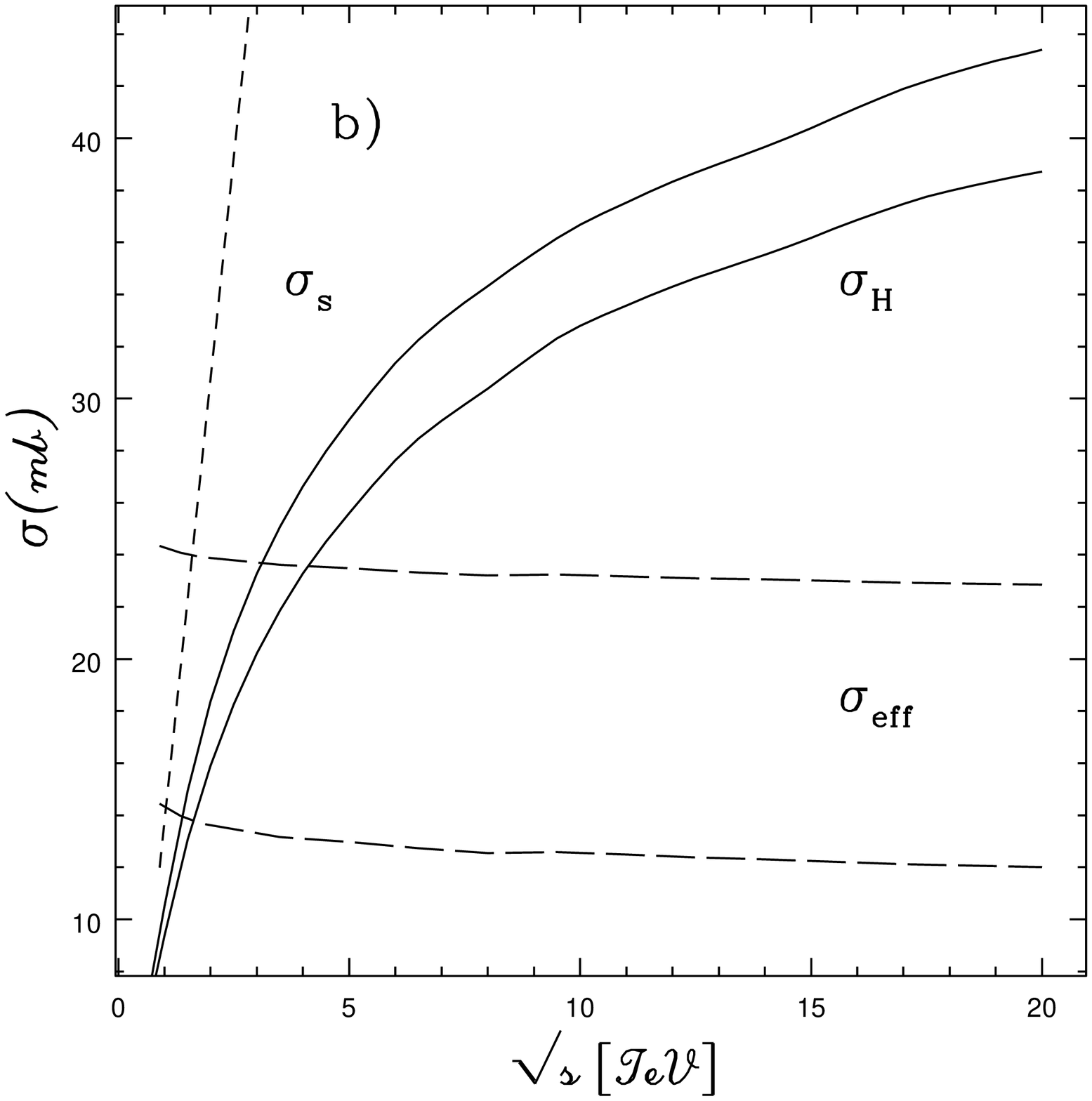}  
}
\vspace{1 cm}
\caption[ ]{ $\sigma_S$, $\sigma_H$ and $\sigma_{eff}$ when taking for $f({\bf r})$ 
a sphere {\it a}) or a gaussian {\it b}). The lower curves for $\sigma_H$ 
and $\sigma_{eff}$ refer to the 
correlated many-body parton distribution, the higher curves to the uncorrelated
ones.}
\label{correl}
\end{figure}

The values of $\sigma_H$ and of $\sigma_{eff}$, derived from the correlated parton distribution described in
the previous paragraph, are plotted in fig.\ref{correl}, where two different analytic
expressions for $f({\bf b})$ are considered. 
In fig.\ref{correl}{\it a} $f({\bf b})$
is the projection of a sharp-edged sphere in transverse space, while in fig.\ref{correl}{\it
b} 
$f({\bf b})$ is a gaussian. 
In each figure we draw the inclusive
cross section $\sigma_S$, which is the fast growing short-dashed curve, the hard
cross section $\sigma_H$, continuous curves, and the effective cross section
$\sigma_{eff}$, long-dashed curves. Both for $\sigma_H$
and for $\sigma_{eff}$ we draw two different curves. The higher curve
refers to the case 
where the gluon and sea distributions
are kept fixed and equal to the average configuration, 
irrespectively of the configuration
of the valence, while the lower curve refers to the case where the 
configuration of 
gluons and sea are correlated in radius with the configuration of the valence, 
as described in the previous paragraph. The lower threshold for
the transverse momentum of the produced jets has been put equal to $5$GeV.
In the correlated case, when
considering a sphere for $f({\bf b})$, we choose for  
the radius of the sphere the value of $B_D$. The RMS of the radius
of the sphere, averaged with the probability distribution of the valence, 
is then $\sqrt{\langle B_D^2\rangle}=0.66$fm, namely it is equal to 
$\sqrt{2/3\langle r^2\rangle}$, with $\sqrt{\langle r^2\rangle}=0.81$fm, the RMS 
proton charge radius. In the gaussian case we fix the size of the distribution
by requiring the same RMS value for the radius of the distribution as in the 
case of the sphere. As one may see in fig.\ref{correl}, while the value of
$\sigma_{eff}$ is too large when the distribution of gluons and sea quarks
is kept fixed, the value of $\sigma_{eff}$ is sizably reduced when it is 
correlated with the distribution of the valence quarks.

With our choice the value of the effective cross section turns out of the right
size. Obviously the result could have been significantly different with a 
different,
but still plausible choice of the (correlated) radius of the gluon and sea 
distributions. We do not claim therefore that the simplest mechanism discussed
here is the only solution to the problem posed by the 
smallness of the observed value of $\sigma_{eff}$.
Rather we point out that the source of correlations discussed here on one
side is a minimal modification to the uncorrelated many-body parton distribution
and, on the other, it looks as 
a natural possibility, which could explain a substantial amount of the 
difference between the observed value of $\sigma_{eff}$ and the 
result of the uncorrelated calculation.

As shown in fig.\ref{correl}, our simplest model, while reducing sizeably
the expectation
for $\sigma_{eff}$, does not modify dramatically the
expectation for $\sigma_H$. It is therefore worth wile making predictions
for other possible observables, in order to have an independent
indication on the actual model. We
have then estimated the triple parton scattering cross section.
The triple scattering
cross section, being proportional to $\sigma_S^3$, introduces a new 
dimensional quantity other than the effective cross section $\sigma_{eff}$.
In the uncorrelated case the new dimensional quantity has a given value,
proportional to $\sigma_{eff}^2$ and the proportionality factor depends on the
actual form of $f({\bf b})$. Also in the correlated case one may however write
\begin{equation}
\sigma_T=\frac{1}{3!}\frac{\sigma_S^3}{\tau\sigma_{eff}^2}
\end{equation}
The observation of the triple scattering parton process allows therefore one 
to measure the dimensionless quantity $\tau$, which, as $\sigma_{eff}$
is a well defined quantity, related to the geometrical properties of the 
interacting hadrons. The actual expectations for the $\tau$-factor according
to the model discussed here are shown in the table. 

\section{Conclusions}

The observation of multiple parton collisions allows the measure of a whole
set of quantities which characterize the interaction and that 
are directly connected
to the geometrical properties of the hadron structure. The first indication
in this direction is the measure of the effective cross section of double
parton collisions performed by CDF. The measured value of the effective 
cross section rules out the
simplest uncorrelated picture of the many body parton distribution of the
hadron. In this note we have shown that correlating the transverse dimension
of the gluon and sea quark distributions to the transverse dimension of the
valence one can obtain for the effective cross section a value much closer 
to the experimental indication. In order to have the possibility to test the model, 
at least to some extent, we have then
worked out our prediction for 
the factor $\tau$, characterizing the triple parton scattering cross section.

The simplest
expectation for $\sigma_{eff}$ and, more in general, for the scale factors 
of the
different multiple parton collision processes, is that they are cutoff and energy
independent. The situation however is changed when more elaborate structures
are considered. Also in the simplest model discussed here in fact 
$\sigma_{eff}$
shows a slight energy and cutoff dependence. The origin
is the following: In our case partons are organized in two different structures
in transverse space, on one side one has valence quarks and on the other gluons and
sea quarks. When changing the energy or the lower cutoff in $p_t$ one is
varying the relative number of interacting sea quarks and gluons with respect to 
the valence quarks. This variation is then reflected in the relative weight
of the rate of collisions of partons which have a different distributions in 
transverse space
and, as a consequence, $\sigma_{eff}$ is slightly modified.

\vskip.25in
{\bf Acknowledgements}

\vskip.25in

This work was partially supported by the Italian Ministry of University and of
Scientific and Technological Research by means of the Fondi per la Ricerca
scientifica - Universit\`a di Trieste.
\begin{table}
 \caption{Factor $\tau$ for the triple parton scattering process}
 \label{tab}
  \begin{tabular}{cccc}
  Sphere (no corr.) & Gauss (no corr.) & Sphere (with corr.) & Gauss (with
  corr.) \\
   \tableline
 0.78        & 0.74      & 0.5     & 0.46   
  \end{tabular}
\end{table}


\end{document}